\documentclass[prd,floats,floatfix,nofootinbib]{revtex4}
\usepackage[dvips]{graphicx}
\usepackage{graphics}
\usepackage{dcolumn}
\usepackage{amssymb}
\usepackage{bm}
\usepackage{amsmath}
\usepackage{color}
\usepackage{subfigure}
\usepackage{epstopdf}

\begin{document}

\title{Scalar Perturbations in Nonsingular Universes from Interacting Vacuum}

\author{Filipe Cattete Alves\footnote{filipe.ca2012@hotmail.com}, Rodrigo Maier\footnote{rodrigo.maier@uerj.br}} 

\affiliation{
Departamento de F\'isica Te\'orica, Instituto de F\'isica, Universidade do Estado do Rio de Janeiro,\\
Rua S\~ao Francisco Xavier 524, Maracan\~a,\\
CEP20550-900, Rio de Janeiro, Brazil
}

\date{\today}

\begin{abstract}
In this paper we examine the stability of scalar perturbations in nonsingular models which emerge from an interacting vacuum component. The analysis developed in this paper relies on two phenomenological choices for the energy exchange between a nonrelativistic fluid and a vacuum component. In both scenarios it can be shown that closed models may furnish nonsingular orbits of physical interest in phase space once a decelerated past era is connected to a graceful exit 
to late-time acceleration. Regarding such configurations as background spacetimes we introduce
scalar perturbations in order to examine the stability of these models in a high energy domain.
We explicitly show that the vacuum perturbation is not an independent variable and diverges as
dynamics approaches the bounce. This feature assigns a rather unstable signature to the dynamics
making the choices for the energy transfer ill defined at least for nonsingular configurations
at the bounce scale.
\end{abstract}

%
% Uncomment for keywords
%\vspace{2pc}
%\noindent{\it Keywords}: XXXXXX, YYYYYYYY, ZZZZZZZZZ
%
% Uncomment for Submitted to journal title message
%\submitto{\JPA}
%
% Uncomment if a separate title page is required
%\maketitle
% 
% For two-column output uncomment the next line and choose [10pt] rather than [12pt] in the \documentclass declaration
%\ioptwocol
%
\maketitle

\section{Introduction}
\label{sec:intro}

The development of astronomical data in the last decades has given a huge support
to favour the $\Lambda$CDM model as the most successful framework to describe our observable Universe. The structure of CMB together with large-scale structure data, the abundance of light elements and late-time acceleration are fundamental
milestones in which modern cosmology relies on (see \cite{wein,amendola} and references therein). However, despite its success it is well known amid cosmologists that General Relativity must be properly corrected or even replaced by a new theory
when sufficiently high energy scales are considered. 
In fact, evolving back the dynamics using Einstein field equations we see that our Universe is driven towards an initial singularity where the classical regime is no longer valid.
In this vein, effective models within the context of modified gravity have been
developed in order to circumvent the initial singularity problem\cite{Novello:2008ra,Maier:2013hr,Maier:2013yh,Bruni:2021msx}.

Apart from the initial singularity, the cosmological constant problem is another issue
which has puzzled cosmologists nowadays. Although the cosmological constant seems to be the simplest candidate 
to trigger late-time acceleration, it suffers from a crucial difficulty when one tries to 
put up
its observed value with theoretical calculations based on quantum field theory\cite{Weinberg:1988cp}.

Also motivated by quantum field
theory considerations\cite{Lima:2013dmf,Moreno-Pulido:2020anb,Moreno-Pulido:2023ryo,SolaPeracaula:2023swx} the possibility of an
interacting vacuum component has been a subject of considerable interest. In the context of black
hole formation for instance, it has been shown\cite{Maier:2020bgm} that a nonsingular collapse
of barotropic perfect fluids
may give rise to Reissner-Nordstr\"om-de Sitter black holes. Yukawa black holes
may also be obtained for an appropriate choice of the energy exchange between
the electromagnetic Maxwell field and a vacuum component\cite{Maier:2021jxv}.
From the cosmological point of view on the other
hand, it has been shown that tensions between
different observational data may be alleviated once an interacting dark energy
component is considered\cite{Salvatelli:2014zta,Wang:2015wga,Zhao:2017cud,Kumar:2017dnp,Wang:2013qy,Martinelli:2019dau}.

The study of structure formation in nonsingular cosmological models is a fundamental task
once a regular bounce may discard the inflationary paradigm by solving
the horizon and flatness problems and justify the power
spectrum of primordial cosmological perturbations inferred by observations\cite{Peter:2008qz,Maier:2011yy,Maier:2013hr}.
In fact, in \cite{Wands:1998yp,Allen:2004vz} it has been shown that curvature perturbations may furnish an
almost scale invariant spectrum if a nonrelativistic fluid dominates the contracting phase
and perturbations
are generated by quantum vacuum fluctuations. We should mention however that although the Bardeen potential\cite{Bardeen:1980kt} grows at the bounce -- making a perturbative analysis
ill defined in this domain -- such inconsistent
behaviour can be avoided by a suitable gauge choice\cite{Allen:2004vz}.
A different treatment can be found in \cite{Maier:2013gua} where scalar perturbations were introduced in regular
nonsingular closed braneworld models. In this context, scalar perturbations were shown to be bounded and sufficiently small for restrictive initial conditions and a proper domain of parameters.

In this paper we examine the stability of nonsingular models due to
an interacting vacuum component. Our analysis is based on two different 
phenomenological choices for the energy transfer between a nonrelativistic fluid
and a vacuum component\cite{Bruni:2021msx}. In Section \ref{sec2} we present the two nonsingular background models
in which our analysis relies on.  In Section \ref{sec3} we introduce scalar pertubations 
and analytically show that the dynamics is unstable for sufficient high energies. 
In Section \ref{sec4} we numerically illustrate such instabilities for the case of closed models considering the development of
the scalar perturbations in a high energy domain as the dynamics approaches the bounce. 
In Section \ref{sec5} we leave our final remarks.

\section{Background Models}
\label{sec2}
We start by considering simple models based on \cite{Bruni:2021msx} in which the initial singularity can be avoided
due to a matter-vacuum energy transfer. 
To this end, let us consider Einstein field equations
\begin{eqnarray}
\label{ein}
G_{\mu\nu}=\kappa^2(V g_{\mu\nu}+T_{\mu\nu}),    
\end{eqnarray}
where $G_{\mu\nu}$ is the Einstein tensor, $\kappa^2\equiv 8\pi G$ and $V$ is the vacuum component. We regard $T_{\mu\nu}$ as the energy-momentum tensor which includes a nonrelativistic fluid with an interacting energy density and $4$-velocity denoted by
%
%\begin{eqnarray}
%T_{\mu\nu}=\rho_I u_\mu u_\nu + {^{(\gamma)}}{T}_{\mu\nu},    
%\end{eqnarray}
%
$\rho_I$ and $u^\mu$, respectively.
%and ${^{(\gamma)}}{T}_{\mu\nu}$ is the noninteracting part of the %energy-momentum tensor
%given by a radiation component.
The interaction between the nonrelativistic fluid and the vacuum component is dictated
by an energy-momentum transfer $4$-vector $Q_\nu$ so that the Bianchi identities furnish
\begin{eqnarray}
\label{eq1}
&&\nabla_\nu V=-Q_\nu\\
\label{eq11}
&&\nabla_\mu (\rho_I u^\mu u_\nu)=Q_\nu.
\end{eqnarray}

In this paper we adhere to the so-called geodesic CDM scenario\cite{Martinelli:2019dau}
in which
\begin{eqnarray}
\label{eq2}
Q^\nu=Q u^\nu,    
\end{eqnarray}
where $Q$ denotes an energy flow. In particular, we are interested in 
nonsingular models\cite{Bruni:2021msx} which emerge from
two different covariant choices for $Q$, namely,
\begin{eqnarray}
\label{eq3}
&&Q_1=\xi (V_\Lambda-V) \nabla_\mu u^\mu,\\
\label{eq31}
&&Q_2=-\chi(1-V/V_\Lambda)\rho_I \nabla_\mu u^\mu.
\end{eqnarray}
In the above, $V_\Lambda$ is a constant parameter which plays a role analogous to the cosmological constant while $\xi$ and $\chi$ are coupling parameters. 

Energy exchanges between a nonrelativistic fluid and a vacuum component
have been considered in the context of running vacuum models (see 
\cite{Lima:2013dmf,Moreno-Pulido:2020anb,Moreno-Pulido:2023ryo,SolaPeracaula:2023swx} and references therein). In this scenario the vacuum component is regarded as a dynamical variable and its 
fluctuation -- which depends on contributions from the quantized matter
fields, namely, bosons and fermions -- is evaluated from quantum field calculations.
Based on this microphysical motivation one should expect that such fluctuation
depends on the Hubble rate $H$ squared. 
%That is,
%
%\begin{eqnarray}
%\label{rvm}
%V_I=\Lambda+\sigma H^2,    
%\end{eqnarray}
%
%where $\Lambda$ is the usual cosmological constant.
It is worth noting however that
the $4$-vector $Q_\nu$ thus obtained from this assumption is observer dependent once it breaks general covariance. Following a different path, pure phenomenological proposals\cite{Wang:2015wga,Zhao:2017cud,Kumar:2017dnp,Wang:2013qy,Valiviita:2008iv,Gavela:2009cy,Costa:2016tpb} for the $4$-vector $Q_\nu$ -- closely related
to running vacuum models -- have been considered. In particular, 
it has been shown that choices 
like (\ref{eq3}) may reduce 
tensions among $H_0$, CMB and cosmic shear/redshift-space distortions measurements\cite{Salvatelli:2014zta,DiValentino:2019ffd,Kaeonikhom:2022ahf}.
Regarding high energy configurations it has also been shown that phenomenological choices
like (\ref{eq3})--(\ref{eq31}) lead to nonsingular models\cite{Bruni:2021msx,Maier:2020bgm}.
In the present paper we intend to examine the stability of such nonsingular models which arise from 
(\ref{eq3})--(\ref{eq31}) in a high energy domain.

In the case of a FLRW metric in conformal time, Einstein field equations
read
\begin{eqnarray}
\label{fi1}
&&{\cal H}^2+k=\frac{\kappa^2}{3}(\rho+\rho_I+V)a^2,\\
\label{fi2}
&&{\cal H}^\prime+\frac{{\cal H}^2}{2}+\frac{k}{2}=\frac{\kappa^2}{2}a^2(V-p),
\end{eqnarray}
where $\rho$ and $p$ stand for the effective energy density and pressure of noninteracting perfect fluids, ${\cal H}\equiv a^\prime/a$ and primes denote differentiation with respect to conformal time.

We start our analysis considering the case $Q=Q_1$.
The substitution of (\ref{eq2})--(\ref{eq3}) in (\ref{eq1})--(\ref{eq11}) furnishes
\begin{eqnarray}
\label{vsol1}
&&V=V_\Lambda+\lambda a^{3\xi},\\
\label{rsol1}
&&\rho_I=\frac{E_d}{a^3}-\frac{\lambda\xi}{1+\xi} a^{3\xi},
\end{eqnarray}
where $\lambda$ and $E_d$ are constants of integration. In (\ref{rsol1}) we identify the first term on the RHS as the conventional nonrelativistic component of the dust fluid density for 
$E_d >0$. 

As in \cite{Bruni:2021msx}, one may also include a noninteracting radiation component. Regarding ${\rho}_\gamma$ as its energy density we gain
\begin{eqnarray}
{\rho}_\gamma=\frac{E_\gamma}{a^4},    
\end{eqnarray}
where $E_\gamma$ is a constant.
In this case the first integral (\ref{fi1}) turns into
\begin{eqnarray}
\label{eqfn}
{\cal H}^2+k=\frac{\kappa^2}{3}\Big(\frac{E_d}{a^3}+\frac{E_\gamma}{a^4}+V_\Lambda+\frac{\lambda a^{3\xi}}{1+\xi}\Big)a^2.
\end{eqnarray}

In order to fix a proper numerical value for the coupling constant $\xi$
we take into account recent results
\cite{Salvatelli:2014zta,DiValentino:2019ffd} in which a late-time interaction of the type (\ref{eq3}) with a negative coefficient $\xi$ (of
the order of $-1$) is favoured by current cosmological measurements as CMB anisotropies, supernovae Ia and redshift space distortions. On the other hand one may easily notice that the further restrictions
\begin{eqnarray}
\label{c1}
\xi=-\frac{4}{3}~~{\rm and}~~3\lambda>E_\gamma,    
\end{eqnarray}
or
\begin{eqnarray}
\label{c2}
\xi<-\frac{4}{3}~~{\rm and}~~\lambda>0,   
\end{eqnarray}
provide 
nonsingular models
regardless of spatial curvature. 
That is, once (\ref{c1}) or (\ref{c2}) are satisfied, our model describes an universe which undergoes through an infinite past contracting phase before it starts to expand after the scale factor reaches a minimum value.
To simplify our analysis in this paper we shall restrict ourselves to the case
(\ref{c1}) so that nonsingular models may be sustained by a 
coupling coefficient $\xi$ (of the order of $-1$) favoured by current observations.
In this case (\ref{eqfn}) reads
\begin{eqnarray}
a^{\prime 2}+U(a)={\cal E},
\end{eqnarray}
\begin{figure}
\begin{center}
\includegraphics[width=7cm,height=5cm]{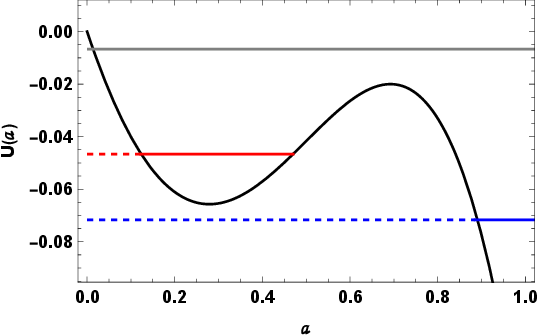}
\caption{The potential $U$ (solid black curve) considering the parameters $k=\kappa=1$, $E_d=1.5$
and $V_\Lambda=2$. The red horizontal line above denotes the energy of one cyclic orbit
(of region I with ${\cal E}\simeq-0.046$)
inside the potential well. 
Gray and blue horizontal lines (with ${\cal E}\simeq-0.006$ and ${\cal E}\simeq-0.071$, respectively) are connected to energies of one bouncing
orbits of regions II and III, respectively. 
The nonsingular orbit with ${\cal E}\simeq-0.006$ is an example of physical interest since it contains a decelerated past era connected to an exit to late-time acceleration.}
\label{figU}
\end{center}
\end{figure}
where
\begin{eqnarray}
U(a)= ka^2-\frac{\kappa^2}{3}\Big({E_d}{a}+V_\Lambda  a^4\Big),
\end{eqnarray}
and ${\cal E}={\kappa^2}({E_\gamma/3-\lambda})$. A straightforward analysis of the potential $U$
shows that it admits at most two local extrema -- one local maximum and one local minimum -- which are connected to fixed points in phase space. In fact, as shown in \cite{Bruni:2021msx} a proper domain
of parameters divide the phase space in three distinct regions -- namely regions I, II and III --
of cyclic and one bouncing orbits. The domain of cyclic universes is referred as region I and contains orbits which oscillate inside a potential well. Nonsingular orbits in region II are of physical interest since they contain a decelerated past era connected to an exit to late-time acceleration. One bouncing orbits can also be found in region III.
In Fig. \ref{figU}
we show the potential $U$ (solid black curve) considering the parameters $k=\kappa=1$, $E_d=1.5$
and $V_\Lambda=2$. Examples of orbits in regions I, II and III are also depicted in this panel.
%It is worth mention that the choice (\ref{c1}) also finds other theoretical applications such 
%as nonsingular black hole formation from dust collapse\cite{Maier:2020bgm}.
It is worth to point out that the results 
shown above are qualitatively valid for other values of 
$\xi$ of similar order once (\ref{c2}) is satisfied, or the second part of (\ref{c1}) if $\xi>-4/3$.

To proceed we now consider the case $Q=Q_2$.
The substitution of (\ref{eq2})--(\ref{eq31}) in (\ref{eq1})--(\ref{eq11}) furnishes
\begin{eqnarray}
\label{eqv14}
&&V^\prime=3\chi(1-V/V_\Lambda)\rho_I{\cal H},\\
&&\rho_I^\prime=-3[1+\chi(1-V/V_\Lambda)]\rho_I{\cal H}.
\end{eqnarray}
From the above one may eliminate ${\cal H}$ so that we end up with
\begin{eqnarray}
\label{diff}
V^\prime+\frac{\chi(1-V/V_\Lambda)}{1+\chi(1- V/V_\Lambda)}\rho_I^\prime=0.    
\end{eqnarray}
It is easy to show that (\ref{diff}) has a first integral given by
\begin{eqnarray}
\label{finl}
\rho_I=\rho_0+(V_\Lambda-V)+\frac{V_\Lambda}{2\chi}\ln{[(V_\Lambda-V)^2]}, 
\end{eqnarray}
where $\rho_0$ is a constant of integration. 
From (\ref{finl}) we note that the number of degrees of freedom is reduced in phase space
so that a $3$-dimensional dynamical system may be obtained. 

To simplify our analysis, in the following we shall assume that the nonrelativistic interacting fluid is the sole component which contributes to the energy momentum tensor so that 
$\rho=p=0$ in (\ref{fi1})--(\ref{fi2}).
In this case, by substituting (\ref{finl}) in (\ref{fi2}) and (\ref{eqv14})
-- with the use of (\ref{fi1}) -- we obtain the following dynamical system
\begin{eqnarray}
&&a^\prime={\cal H}a,\\
\label{eqv}
&&V^\prime=3\chi{\cal H}(1-V/V_\Lambda)\Big\{(\rho_0+V_\Lambda-V)+\frac{V_\Lambda}{2\chi}\ln{[(V_\Lambda-V)^2]}\Big\},\\
\label{eqh}
&&{\cal H}^\prime=-\frac{\kappa^2a^2}{6}\Big\{\rho_0+V_\Lambda-3V+\frac{V_\Lambda}{2\chi}\ln{[(V_\Lambda-V)]^2}\Big\}.
\end{eqnarray}

\begin{figure}
\begin{center}
\includegraphics[width=7cm,height=5cm]{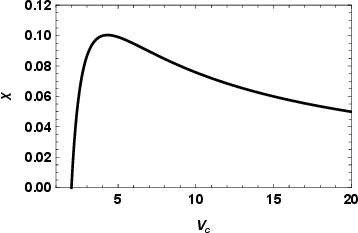}
\caption{The coupling parameter $\chi$ as a function of $V_c$ according to (\ref{chi1}). From this figure we see that fixing a constant value for $\chi$ there is at most two fixed points for the proper domain $0< \chi \lesssim 0.1$.}
\label{fig1}
\end{center}
\end{figure}

Following a procedure analogous to that one developed in \cite{Bruni:2021msx} we shall consider bouncing solutions
by probing for saddle and center fixed points with ${\cal H}=0$. To this end, according to (\ref{eqh}) such fixed points must satisfy
\begin{eqnarray}
\label{eqf}
2\chi(V_\Lambda+\rho_0-3V_c)+V_\Lambda\ln[(V_\Lambda-V_c)^2]=0,
\end{eqnarray}
where $V_c$ stands for the numerical value $V$ evaluated at the fixed point.
In order to furnish a numerical illustration let us consider the case $\rho_0=0$ and $V_\Lambda=1$.
In this case (\ref{eqf}) shows that there is at most two fixed points for a proper domain of $\chi$ once the following relation 
\begin{eqnarray}
\label{chi1}
\chi=\frac{\ln[(1-V_c)^2]}{2(3V_c-1)}    
\end{eqnarray}
is satisfied. In fact, in the domain $0 < \chi \lesssim 0.1$ there are two possible values for $V_c$ -- say $V_{c-}$ and $V_{c+}$, with $V_{c-}<V_{c+}$ -- as illustrated in Fig. \ref{fig1}. It can be shown\cite{Bruni:2021msx} that $V_{c-}$ corresponds to a saddle
while $V_{c+}$ is connected to a center fixed point.
In fact, defining $\Psi_i$ as the $2$-vector
\begin{eqnarray}
\Psi_i=\begin{pmatrix} V-V_c \\ {\cal H}-{\cal H}_c \end{pmatrix}    
\end{eqnarray}
one may expand (\ref{eqv})--(\ref{eqh}) in a neighbourhood of the fixed points as
\begin{eqnarray}
\Psi^\prime_i=L_{ij}\Psi_j    
\end{eqnarray}
where $L_{ij}\equiv\partial\Psi_i^\prime/\partial\Psi_j$ is evaluated at the fixed points.
A straightforward evaluation then furnishes that the eigenvalues of $L_{ij}$ are given by
\begin{eqnarray}
\lambda_{\pm}=\pm\kappa a_c \sqrt{V_c\Big[1+3\chi\Big(1-{V_c}\Big)\Big]}.  
\end{eqnarray}
In the above, $a_c$ stands for the numerical value $a$ evaluated at the fixed point.
As an example, for $\chi=0.08$ and $\kappa=1$, we obtain
that $\lambda_{\pm}(V_{c-})\simeq \pm1.26 a_c$ while
$\lambda_{\pm}(V_{c+})\simeq \pm2.87 {\rm i} a_c$.
Therefore $V_{c-}$ is a saddle while $V_{c+}$ is a center fixed point. 

\begin{figure}
\begin{center}
\includegraphics[width=7cm,height=5cm]{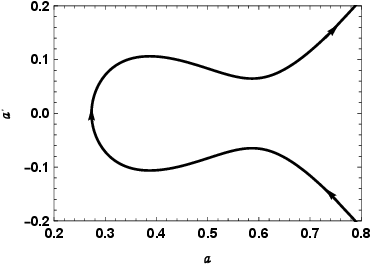}
\caption{A nonsingular orbit of physical interest. In this panel we see that during its expanding phase (${\cal H}>0$) $a^\prime$ decreases in the domain
$0.34\lesssim a \lesssim 0.58$ illustrating a decelerated era. For $0.58 \lesssim a$
the orbit escapes towards late-time acceleration.
In this example we fixed the parameters $\chi=0.08$ and $\kappa=1$.}
\label{fig2}
\end{center}
\end{figure}

As shown in \cite{Bruni:2021msx} the general picture of the phase
space shows perpetually bouncing solutions in a finite neighbourhood of the center fixed point. This neighbourhood is bounded by a separatrix -- an homoclinic orbit with de Sitter attractors -- which emerges from the saddle fixed point. Higher energy nonsingular orbits of physical interest lay outside the separatrix and exhibit a decelerated past era during its expanding phase, followed by a graceful exit 
to late-time acceleration. In Fig. \ref{fig2} we show an example of such orbit.
It is worth to mention that this orbit exhibits the same general features as the example depicted in Fig. \ref{figU} by the gray horizontal line.

\section{Scalar Perturbations}
\label{sec3}
In order to introduce metric scalar perturbations $\phi$ and $\psi$, let us consider 
the perturbed FLRW line element in the longitudinal gauge\cite{Mukhanov:1990me}
\begin{eqnarray}
\label{met}
ds^2=a^2(\eta)[(1+2\phi)d\eta^2-(1-2\psi)\gamma_{ij}d\zeta^i d\zeta^j],    
\end{eqnarray}
where $\zeta^i$ stand for comoving coordinates $(r, \theta, \varphi)$ and 
\begin{eqnarray}
\gamma_{ij}={\rm diag}\Big(\frac{1}{1-kr^2}, r^2, r^2\sin^2\theta\Big).    
\end{eqnarray}

For a general perfect fluid with 
energy-momentum tensor
\begin{eqnarray}
T^{\mu\nu}=(\rho+p)u^\mu u^\nu-pg^{\mu\nu},    
\end{eqnarray}
we introduce hydrodynamical perturbations as
\begin{eqnarray}
\delta {T}^{\mu}_{~\nu}=(\delta \rho+\delta p)u^\mu u_\nu-\delta p\delta^\mu_{~\nu}+\delta q^\mu u_{\nu}+\delta q_{\nu}u^\mu.
\end{eqnarray}
In the above $\delta q_\mu$ is connected to a heat flux.
Such perturbations are pure spatial and we define them as 
\begin{eqnarray}
\label{deltaq}
\delta q_i= -a {^{(3)}}\nabla_i \delta q
\end{eqnarray}
where $\delta q$ is a scalar function and ${^{(3)}}\nabla_i$ stands for the
covariant derivative built with the spatial metric $\gamma_{ij}$.
We should stress that in the present context we shall neglect 
any sort of anisotropic pressure perturbation in order to simplify our analysis.
In fact, the absence of such nondiagonal perturbations 
lead to the 
significant simplification $\phi\equiv\psi$.

From (\ref{ein}) the perturbed Einstein field
equations can be written as
\begin{eqnarray}
\label{eqp0}
&&\Delta \phi-3{\cal H}(\phi^\prime+{\cal H} \phi)+3k\phi=\frac{\kappa^2 a^2}{2}(\delta \rho+\delta V),\\
\label{pert2}
&&\phi^\prime+{\cal H \phi}=-\frac{\kappa^2 a^2}{2}\delta q,\\
\label{pert3}
&&(2{\cal H}^\prime+{\cal H}^2)\phi+3{\cal H}\phi^\prime+\phi^{\prime\prime}-k\phi=\frac{\kappa^2 a^2}{2}\Big(\delta p-\delta V\Big),
\end{eqnarray}
where $\Delta\equiv \gamma^{ij}{^{(3)}}\nabla_i{^{(3)}}\nabla_j$.

The remaining equations comes from the perturbed part of the conservation equations
\begin{eqnarray}
\label{pp}
\nabla_\mu(T^\mu_{~\nu}+V\delta^\mu_{~\nu})\equiv 0.    
\end{eqnarray}
For $\nu=0$ we obtain,
\begin{eqnarray}
\label{p1}
&&\delta V^\prime=-\delta Q_0,\\
\label{p2}
&&\delta \rho^\prime+3{\cal H}(\delta\rho+\delta p)-3\phi^\prime(\rho+p)-a^2\Delta\delta q=\delta Q_0.~~~~
\end{eqnarray}
The spatial sector on the other hand furnishes,
\begin{eqnarray}
\label{p3}
&&{^{(3)}}\nabla_i\delta V=-\delta Q_i,\\
\label{p4}
&&{^{(3)}}\nabla_i\Big[\delta q^\prime+4{\cal H}\delta q+(\rho+p)\phi+\delta p\Big]=-\delta Q_i.
\end{eqnarray}
At this stage we should mention that the relation 
\begin{eqnarray}
\label{rel}
\delta Q_i={^{(3)}}\nabla_i \bar{Q},    
\end{eqnarray}
where
$\bar{Q}$ is an homogeneous function, is a sufficient condition to provide a consistent set of
equations for scalar perturbations. In fact, in this case it can be easily shown that (\ref{pert3}) can be obtained by taking the derivative with respect to conformal time of (\ref{pert2}). In the following we shall see that our model satisfy (\ref{rel}) once we take into account the choices (\ref{eq3})--(\ref{eq31}).
It is worth noting that 
the results obtained below do not depend on a gauge choice once
the metric perturbation $\phi$
in the equations of motion (\ref{eqp0})--(\ref{pert3}) coincide with the gauge-invariant variable as shown in \cite{Mukhanov:1990me}.

Let us then consider the case $Q=Q_1$.
As apart from nonrelativistic matter a radiation component is also considered in this case, in previous equations we assume that
\begin{eqnarray}
\nonumber
&&\rho\rightarrow \rho_I+\rho_\gamma,\\
\nonumber
&&p\rightarrow \frac{1}{3}\rho_\gamma,\\
\nonumber
&&\delta\rho\rightarrow\delta\rho_I+\delta\rho_\gamma,\\
\nonumber
&&\delta p \rightarrow \frac{1}{3}\delta\rho_\gamma,\\
\nonumber
&&\delta q \rightarrow \delta q_I+\delta q^{(\gamma)},
\end{eqnarray}
where we have fixed the equation of state $\delta p = w \delta \rho$
for hydrodynamical perturbations as well.

As the radiation component does not induce and energy-momentum transfer, from (\ref{p2}) and (\ref{p4}) we obtain:
\begin{eqnarray}
&&\delta\rho_\gamma^\prime=-4{\cal H}\delta\rho_\gamma+4\phi^\prime\rho_\gamma+a^2\Delta \delta q^{(\gamma)},\\
&&\delta q^{\prime(\gamma)}=-4{\cal H}\delta q^{(\gamma)}-\frac{4}{3}\rho_\gamma\phi-\frac{1}{3}\delta{\rho_\gamma}.
\end{eqnarray}

To complete our analysis a proper evaluation of the perturbed energy transfer $4$-vector $Q_\nu$ in (\ref{eq2})--(\ref{eq3}) must be made in order to obtain the evolution equations of scalar perturbations. Considering the choice (\ref{eq3}), we are then led to
\begin{eqnarray}
\label{deltaQ}
\delta Q_\nu = -\xi  [{^{(0)}}\nabla_\mu u^\mu]u_\nu\delta V+\xi(V_\Lambda-V)\Big\{\delta (\nabla_\alpha u^\alpha)u_\nu
+[{^{(0)}}\nabla_\mu u^\mu]\delta u_\nu\Big\}.
\end{eqnarray}
From the above we notice that the perturbed $4$-vector $\delta Q_\nu$ induces a momentum exchange between vacuum and the nonrelativistic matter, apart from an energy transfer. 
Given the definition (\ref{deltaq}) it can be easily shown
that the heat flux perturbation for the interacting matter can be written as
\begin{eqnarray}
\label{deltaq1}
 {^{(3)}}\nabla_i \delta q_I=-\Big(\frac{\rho_I}{a}\Big)\delta u_i  
\end{eqnarray}
according to \cite{Mukhanov:1990me}. Therefore, the substitution of 
(\ref{deltaq1}) in (\ref{deltaQ}) furnishes
\begin{eqnarray}
\nonumber
\delta Q_0 = -{\xi} \Big\{ 3{\cal H}\delta V
+(V_\Lambda-V)\\
\label{dq0}
\times\Big[{-2{\kappa^2a^2}(\delta q_I+\delta q^{(\gamma)})+{2{\cal H}}\phi}
-\Big(\frac{1}{\rho_I}\Big)\Delta \delta q_I\Big]\Big\},
\end{eqnarray}
where we have used (\ref{pert2}). On the other hand, the spatial sector of (\ref{deltaQ}) yields
\begin{eqnarray}
\label{dqi1}
\delta Q_i={^{(3)}}\nabla_i\delta\bar{Q},
\end{eqnarray}
with
\begin{eqnarray}
\label{dqi2}
\delta\bar{Q}=- \frac{3\xi{\cal H}}{\rho_I}(V_\Lambda-V) \delta q_I.    
\end{eqnarray}

The substitution of (\ref{dq0})--(\ref{dqi2}) in the interacting sector of (\ref{p2})--(\ref{p4}) furnishes:
\begin{eqnarray}
&&\delta \rho_I^\prime+3{\cal H}\delta\rho_I-3\phi^\prime\rho_I-a^2\Delta\delta q_I=\delta Q_0,\\
\label{eq38}
&&\delta V=\frac{3\xi{\cal H}}{\rho_I}(V_\Lambda-V) \delta q_I,\\
\label{eq39}
&&\delta q_I^\prime+4{\cal H}\delta q_I+\rho_I\phi=\frac{3\xi{\cal H}}{\rho_I}(V_\Lambda-V) \delta q_I.
\end{eqnarray}
From (\ref{eq38}) and (\ref{eq39}) it can be shown that
\begin{eqnarray}
\nonumber
\delta V^\prime&=&-\Big\{6{\cal H}\rho_I\phi-\delta q_I\Big[\kappa^2a^2(2V-\rho_I-2\rho_\gamma)-6(1-3\xi){\cal H}^2\Big]\Big\}\\
\label{vln}
&\times&\frac{\xi(V_\Lambda-V)}{2\rho_I}
\end{eqnarray}
where we have used (\ref{fi1}) and (\ref{fi2}). From the above one may 
notice that $\delta V$ is not an independent perturbation variable. In fact, subtracting (\ref{vln}) from (\ref{p1}) we obtain
\begin{eqnarray}
\nonumber
&&\delta V=\frac{V_\Lambda-V}{6{\cal H}\rho_I}\Big\{-10{\cal H}\rho_I\phi+4\kappa^2\rho_I a^2 \delta q^{(\gamma)}\\
\label{vp}
&&+\Big[\kappa^2a^2(2V-\rho_I-2\rho_\gamma)-6(1-3\xi){\cal H}^2+{2(2\kappa^2a^2\rho_I+\Delta)}\Big]\delta q_I\Big\}.
\end{eqnarray}
The above result indicates that the dynamics might be highly unstable for sufficient high energy configurations once $\delta V$ seems to diverge as ${\cal H} \rightarrow 0$ at the bounce. In addition, assuming that the scale factor reaches sufficiently small scales  
as we go back in time, $\rho_I$ must vanish at some $\bar{a}=4\lambda/E_d$ before the bounce. Taking into account that 
numerator of (\ref{vp}) does not depend
sole on $\rho_I$ we see that $\delta V$ may
blow up at $\bar{a}$ even before the bounce.

Moving to the case $Q=Q_2$, 
%we notice that as we are neglecting the radiation component, 
in (\ref{eqp0})--(\ref{p4}) we assume
\begin{eqnarray}
\nonumber
\rho\rightarrow \rho_I,~~p\rightarrow 0,~~\delta\rho\rightarrow\delta\rho_I,~~\delta p \rightarrow 0,~~\delta q \rightarrow \delta q_I.
\end{eqnarray}
%
%where we have again fixed the equation of state $\delta p = w \delta \rho$
%for hydrodynamical perturbations.

Considering the choice (\ref{eq31}), we now obtain
\begin{eqnarray}
\nonumber
\delta Q_0 &=&{\chi} \Big\{ \frac{3{\cal H}\rho_I}{V_\Lambda}\delta V\\
\label{dq02}
&-&(1-V/V_\Lambda)\Big[3{\cal H}\delta\rho_I+\rho_I\Big({2{\kappa^2a^2}\delta q_I-2{\cal H}\phi}\Big)
+\Delta \delta q_I\Big]\Big\},
\end{eqnarray}
and
\begin{eqnarray}
\label{dqi12}
\delta Q_i={^{(3)}}\nabla_i\delta\bar{Q},
\end{eqnarray}
with
\begin{eqnarray}
\label{dqi22}
\delta\bar{Q}={3\chi{\cal H}}(1-V/V_\Lambda)\delta q_I.    
\end{eqnarray}

The substitution of (\ref{dq02})--(\ref{dqi22}) in the interacting sector of (\ref{p2})--(\ref{p4}) furnishes:
\begin{eqnarray}
&&\delta \rho_I^\prime+3{\cal H}\delta\rho_I-3\phi^\prime\rho_I-a^2\Delta\delta q_I=\delta Q_0,\\
\label{eq382}
&&\delta V=-3\chi{\cal H}(1-V/V_\Lambda)\delta q_I,\\
\label{eq392}
&&\delta q_I^\prime+4{\cal H}\delta q_I+\rho_I\phi=-3\chi{\cal H}(1-V/V_\Lambda)\delta q_I.
\end{eqnarray}
From (\ref{eq382}) and (\ref{eq392}) it can be shown that
\begin{eqnarray}
\label{vln2}
\delta V^\prime=\frac{\chi}{2}(1-V/V_\Lambda)
\Big\{6{\cal H}\Big[\rho_I\phi+{\cal H}\delta q_I\Big(4+3\chi\Big(1-\frac{V-\rho_I}{V_\Lambda}\Big)\Big)\Big]\\
\nonumber
+\kappa^2a^2\delta q_I(\rho_I-2V)\Big\}
\end{eqnarray}
where we have used (\ref{fi1}) and (\ref{fi2}). From the above one may 
notice again that $\delta V$ is not an independent perturbation variable. Subtracting (\ref{vln2}) from (\ref{p1}) we obtain
\begin{eqnarray}
\label{vp2}
\delta V=\frac{V_\Lambda-V}{6{\cal H}\rho_I}\Big\{\Big[2\Delta+\kappa^2a^2(2V+3\rho_I)\\
\nonumber
-6{\cal H}^2\Big(4+3\chi\Big(1-\frac{V-\rho_I}{V_\Lambda}\Big)\Big)\Big]\delta q_I
+2{\cal H}(3\delta\rho_I-5\rho_I\phi)\Big\}.
\end{eqnarray}
Again, (\ref{vp2}) indicates that for sufficient high energy configurations $\delta V$ diverges as $a\rightarrow \bar{a}$ or ${\cal H} \rightarrow 0$ at the bounce.
%Furthermore, taking into account that 
%numerator of (\ref{vp2}) does not depend
%sole on $\rho_I$, the vacuum perturbation $\delta V$ may
%blow up even before the bounce once the scale factor reaches sufficiently small scales  
%so that $\rho_I$ must vanish at $\bar{a}=4\lambda/E_d$.

\section{Instability in Closed Models}
\label{sec4}

In this section we examine the instability of scalar perturbations in a high energy regime
considering their growth through the bounce in closed models with $k=1$. To this end we
expand the spatial part of scalar perturbations
in terms of the eigenmodes $\zeta_{nlm}\equiv R_{nl}(r)Y_{lm}(\theta, \varphi)$ of the Laplacian operator defined over the $3$-sphere so that
$
\Delta \zeta_{nlm}=(n^2-1)\zeta_{nlm},    
$
where $n$ is a positive integer and $\zeta$ stands for $\phi$, $\delta\rho_I$, $\delta\rho_\gamma$, and so on.
It can be shown that the modes $n=1$ and $n=2$
are not physically relevant since they 
correspond to an homogeneous deformation of the curvature radius
and a displacement of the centre of the $3$-sphere, respectively\cite{Lehoucq:2002wy}.
An analogous treatment can be found in \cite{Maier:2013gua} where
scalar
perturbations are examined through their Fourier modes so that
the differential equations become
decoupled for each mode. In the following we shall drop the mode indices in order to avoid
overcluttered formulae.

\begin{figure*}
\includegraphics[width=7.5cm,height=5cm]{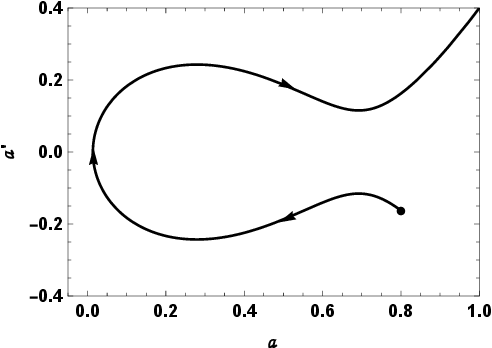}
\includegraphics[width=7cm,height=5cm]{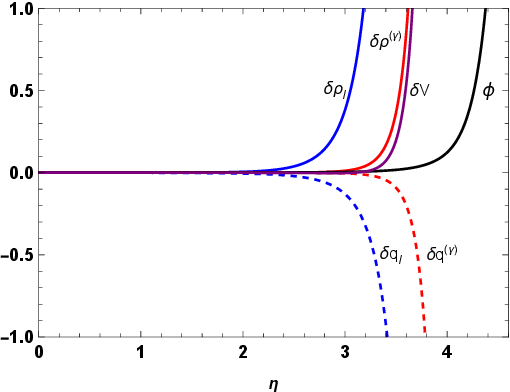}
\caption{A nonsingular orbit of physical interest (left panel) and the evolution of its scalar perturbations (right panel). The background initial conditions illustrated by the bold black dot on the left panel are given by $a_0=0.8$ and $a^\prime_0\simeq -0.162$ -- see text.
The metric and hydrodynamical pertubations evolution in a high energy regime during the contracting phase are shown on the right panel for the mode $n=3$. In this example we show that scalar perturbations diverge before the dynamics reaches the bounce at $\eta\simeq 4.86$. In the above we have fixed initial conditions $\phi_0=10^{-4}$, $\delta\rho_I=\delta\rho^{(\gamma)}=\delta q_I=\delta q^{(\gamma)}=0$.}
\label{fig3}
\end{figure*}

To proceed we construct a general dynamical system 
fed with the background nonsingular solutions presented in Section \ref{sec2}. 
Our particular interest concerns one bouncing solutions 
(as in Fig. 3) in which during its expanding phase (with ${\cal H}>0$) a decelerated era is followed by an exit to late-time acceleration.

For $Q=Q_1$ we define
the following dynamical system
\begin{eqnarray}
&&a^\prime={\cal H}a\\
&&{\cal H}^\prime=-\frac{{\cal H}^2}{2}-\frac{k}{2}+\frac{\kappa^2}{2}a^2\Big(V-\frac{\rho_\gamma}{3}\Big)\\
&&\phi^\prime=-{\cal H}\phi-\frac{\kappa^2a^2}{2}[\delta q_I+\delta q^{(\gamma)}]\\
\nonumber
&&\delta \rho_I^\prime=-3\Big\{{\cal H}\delta\rho_I+\Big[{\cal H}\phi+\frac{\kappa^2a^2}{2}(\delta q_I+\delta q^{(\gamma)})\Big]\rho_I\Big\}\\
&&+a^2(n^2-1)\delta q_I-{\xi} \Big\{ 3{\cal H}\delta V
+(V_\Lambda-V)~~~~~~~~~~~~~~~~~~~~~~\\
&&\times\Big[{-2{\kappa^2a^2}(\delta q_I+\delta q^{(\gamma)})+{2{\cal H}}\phi}
-\Big(\frac{n^2-1}{\rho_I}\Big)\Delta \delta q_I\Big]\Big\},\\
&&\delta q_I^\prime=-4{\cal H}\delta q_I-\rho_I\phi+\frac{3\xi{\cal H}}{\rho_I}(V_\Lambda-V) \delta q_I,\\
\nonumber
&&\delta\rho_\gamma^\prime=-4{\cal H}\delta\rho_\gamma-4\Big\{{\cal H}\phi+\frac{\kappa^2a^2}{2}[\delta q_I+\delta q^{(\gamma)}]\Big\}\rho_\gamma\\
&&~~~~~~~~~~~~~~~~~~~~~~~~~~~~~~~~+a^2(n^2-1) \delta q^{(\gamma)},\\
&&\delta q^{\prime(\gamma)}=-4{\cal H}\delta q^{(\gamma)}-\frac{4}{3}\rho_\gamma\phi-\frac{1}{3}\delta{\rho_\gamma},
\end{eqnarray}
subjected to the constraint (\ref{eqp0}), namely
\begin{eqnarray}
\label{cons}
&&(n^2-1) \phi-3{\cal H}(\phi^\prime+{\cal H} \phi)+3k\phi=\frac{\kappa^2 a^2}{2}(\delta\rho+\delta V).
\end{eqnarray}
In the above, the matter and vacuum energy densities satisfy the background solutions 
(\ref{vsol1})--(\ref{rsol1}).
Furthermore, the vacuum perturbation $\delta V$ can be written in terms of the remaining perturbed variables according to (\ref{vp}). 
\begin{figure*}
\includegraphics[width=7.5cm,height=5cm]{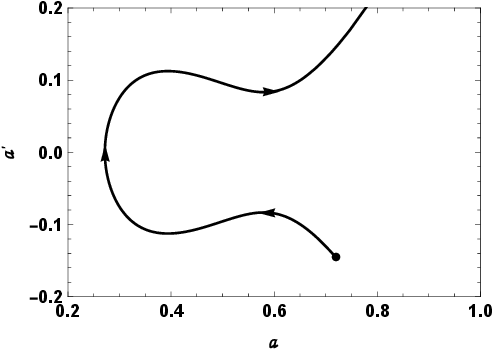}
\includegraphics[width=7cm,height=5cm]{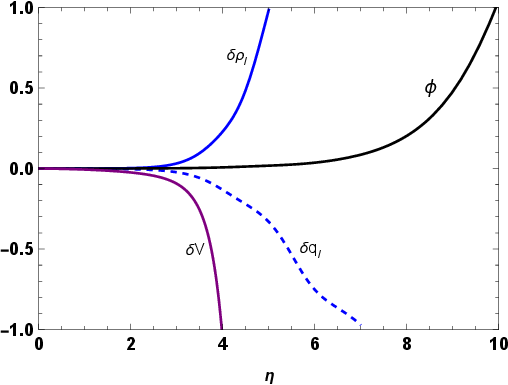}
\caption{A nonsingular orbit of physical interest (left panel) and the evolution of its scalar perturbations (right panel). The background initial conditions illustrated by the bold black dot on the left panel are given by $a_0\simeq 0.71$ and ${\cal H}_0= -0.2$ -- see text.
The metric and hydrodynamical pertubations evolution in a high energy regime during the contracting phase are shown on the right panel for the mode $n=3$. In this example we show that scalar perturbations diverge before the dynamics reaches the bounce at $\eta\simeq 4.90$. In the above we have fixed initial conditions $\phi_0=10^{-4}$, $\delta\rho_I=\delta\rho^{(\gamma)}=\delta q_I=\delta q^{(\gamma)}=0$.}
\label{fig4}
\end{figure*}

Restricting ourselves to the case $\xi=-4/3$ we fix the background parameters as
$V_\Lambda=2, \lambda=E_\gamma=0.01, E_d=1.5$ and $\kappa^2=1$. Fixing the initial condition
$a_0=0.8$, from (\ref{fi1}) we gain ${\cal H}_0\simeq -203$ and we obtain a nonsingular orbit of physical interest since a decelerated past era is connected to a graceful exit 
to late-time acceleration. This orbit is illustrated on the left panel of Fig. \ref{fig3}.
The evolution of the corresponding metric and hydrodynamical perturbations is shown on the right panel of Fig. \ref{fig3}. In this example we show that scalar perturbations diverge before the dynamics reaches the bounce.

For $Q=Q_2$
the following dynamical system can be defined
\begin{eqnarray}
&&a^\prime={\cal H}a\\
\nonumber
&&{\cal H}^\prime=-\frac{\kappa^2a^2}{6}\Big\{\rho_0+V_\Lambda-3V\\
&&~~~~~~~~~~~~~~~~~~~~~+\frac{V_\Lambda}{2\chi}\ln{[(V_\Lambda-V)]^2}\Big\},\\
\nonumber
&&V^\prime=3\chi{\cal H}(1-V/V_\Lambda)\Big\{(\rho_0+V_\Lambda-V)\\
&&~~~~~~~~~~~~~~~~~~~~+\frac{V_\Lambda}{2\chi}\ln{[(V_\Lambda-V)^2]}\Big\},\\
&&\phi^\prime=-{\cal H}\phi-\frac{\kappa^2a^2}{2}\delta q_I\\
\nonumber
&&\delta \rho_I^\prime=-3\Big\{{\cal H}\delta\rho_I+\Big[{\cal H}\phi+\frac{\kappa^2a^2}{2}\delta q_I\Big]\rho_I\Big\}\\
&&+a^2(n^2-1)\delta q_I+{\chi} \Big\{ \frac{3{\cal H}\rho_I}{V_\Lambda}\delta V
-(1-V/V_\Lambda)\\
\nonumber
&&\times\Big[3{\cal H}\delta\rho_I+\rho_I\Big({2{\kappa^2a^2}\delta q_I-2{\cal H}\phi}\Big)
+\Delta \delta q_I\Big]\Big\}\\
&&\delta q_I^\prime=-4{\cal H}\delta q_I-\rho_I\phi-3\chi{\cal H}(1-V/V_\Lambda)\delta q_I,
\end{eqnarray}
again subjected to the constraint (\ref{cons}).
In this case, the vacuum perturbation $\delta V$ can be written in terms of the remaining perturbed variables according to (\ref{vp2}).

Restricting ourselves to the case $\chi=0.08$ we fix the background parameters as
$V_\Lambda=1$ and $\kappa^2=1$. Fixing the initial condition
${\cal H}_0\simeq 0.71$, from (\ref{fi1}) we gain $a_0\simeq 0.71$ and we obtain a nonsingular orbit of physical interest since a decelerated past era is connected to a graceful exit 
to late-time acceleration. This orbit is illustrated on the left panel of Fig. \ref{fig4}.
The evolution of the corresponding metric and hydrodynamical perturbations is shown on the right panel of Fig. \ref{fig4}. In this example we show that scalar perturbations diverge before the dynamics reaches the bounce.

\section{Final Remarks}
\label{sec5}

In this paper we examine the stability of scalar perturbations in nonsingular models which arise from an interacting vacuum component. In the present context we consider two phenomenological choices for the energy exchange between a nonrelativistic fluid and a vacuum component.
In both cases it can be shown that there is a domain in which the universe reaches a minimum volume -- after an infinite past contracting era -- and starts to expand towards a decelerated phase followed by late-time acceleration regime. By introducing scalar perturbations we 
verify that the vacuum perturbation is not an independent variable. Moreover, we analytically show that such perturbation is ill defined once it grows substantially in a high energy domain before diverging at the bounce. We numerically illustrate this behaviour considering the case of closed models. To this end we expand the spatial part of scalar perturbations
in terms of the eigenmodes of the Laplacian operator defined over the $3$-sphere.
In this sense, scalar perturbations are examined through their Fourier modes so that
the differential equations become decoupled for each mode. For the mode $n=3$ we numerically show
(for both choices of the energy transfer)
that all perturbed variables diverge as the dynamics approaches the bounce after an infinite past contracting phase.

We should remark that although choices like (\ref{eq3})--(\ref{eq31}) have been considered in the literature (see \cite{Bruni:2021msx,Kaeonikhom:2022ahf} and references therein),
one should expect that they are ill defined in the context of nonsingular cosmologies once they
are linear functions of the expansion parameter $\nabla_\mu u^\mu$ -- 
 the Hubble rate ${\cal H}$. In fact, let us consider a general energy transfer of the type
\begin{eqnarray}
Q_\nu=Q(t)(\nabla_\mu u^\mu)u_\nu,    
\end{eqnarray}
where $Q(t)$ is a general homogeneous function which does not depend on ${\cal H}$.
From (\ref{p1}), (\ref{p3}) and (\ref{p4}) one may show that at sufficient high energies
\begin{eqnarray}
\delta Q \propto \frac{Q(t)}{\rho_I{\cal H}}\delta q_I,    
\end{eqnarray}
so that $\delta Q$ -- and hence all scalar perturbations -- diverges at the bounce
unless $\delta q_I/{\cal H}$ remains finite.

As a future perspective we intend to better examine different nonsingular models which come
from an interacting vacuum which furnish a consistent analysis of scalar perturbations
through their bounce. Special cases in which $\delta q_I/{\cal H}$ remains finite will
also deserve further examination once they may sustain the choices of energy transfers adopted in this paper.

For the last but not least, it is well known that the cosmological fine-tuning is a major issue
in cosmology nowadays. As this problem turns out to be manifest at the high IR energy domain
-- which the analysis of this paper does not concern -- we do not propose
an alternative mechanism to address it.
We are aware that an interacting dark energy can make the cosmological constant problem
worse in some cases\cite{Marsh:2016ynw}. We intend to address this issue in a further publication.

\section{Acknowledgments}

FCA is supported by CAPES Grant No. $88887.837937/2023-00$.
RM acknowledges financial support from
FAPERJ Grant No. E-$26/010.002481/2019$.

% Bibliography

%% [A] Recommended: using JHEP.bst file
%% \bibliographystyle{JHEP}
%% \bibliography{biblio.bib}

%% or
%% [B] Manual formatting (see below)
%% (i) We suggest to always provide author, title and journal data or doi:
%% in short all the informations that clearly identify a document.
%% (ii) please avoid comments such as "For a review'', "For some examples",
%% "and references therein" or move them in the text. In general, please leave only references in the bibliography and move all
%% accessory text in footnotes.
%% (iii) Also, please have only one work for each \bibitem.

\end{document}